\pdfoutput=1
\documentclass[twocolumn,amssymb,fleqn] {revtex4}
\usepackage{epsfig,amssymb,amsmath,graphicx,subfigure,hyperref  }  
\usepackage{siunitx}
\usepackage{mathrsfs}
\usepackage{array}
\usepackage{lipsum}
\usepackage{color}
\usepackage{multirow}
\usepackage{tabularx}
\usepackage{caption}
\usepackage{bm}

\begin{document}
\captionsetup[figure]{labelfont={bf},name={FIG. },labelsep=period,justification=raggedright}

\title{Tunable phase transition in a dissipative two-spin system: A renormalization group study} 
\author{Xingdong Luo\;}
\email{luo-xingdong@sjtu.edu.cn}
\affiliation{Wilczek Quantum Center and Key Laboratory of Artificial Structures and Quantum Control, School of Physics and Astronomy, Shanghai Jiao Tong University, Shanghai 200240, China} 
\begin{abstract}
We proposed and theoretically studied a model of two separated  spins coupled to a common bosonic bath. 
In our SU(2)-symmetric model, the phase transition point and the stable fixed point representing the nonclassical phase can be continuously tuned by the distance between the two spins. Thus the fixed point collision and annihilation (FPCAA) could be achieved by simply changing the distance. Moreover, we studied in detail  variation of the critical value when the distance is changed. Finally, we also discussed several possible generalizations of our model which may be worth further studying. Our work is essential for  realization of FPCAA in such systems and paves the way for studying related physics such as deconfined criticality and quasiuniversality.
\end{abstract}
\maketitle

\section{Introduction}
 Spin-boson model has been widely studied\cite{vojta2006impurity,kehrein1996spin,bulla2003numerical,winter2009quantum,alvermann2009sparse,zhang2010quantum,cai2014identifying},  due to its importance in many fields of physics, including not only fundamental statistical mechanics\cite{leggett1987dynamics} and quantum dissipation\cite{weiss1993quantum}, but also quantum
optics\cite{walls1994atomic}, nuclear physics\cite{klein1991boson} and quantum chaos\cite{haake1991quantum}. Single-component spin-boson model is a paradigmatic dissipative spin model where the $z$ component of a $S=1/2$ quantum spin is coupled to a single bosonic bath with power law spectral density $ \omega^{1-\delta}$.  Its universality class of phase transition is identical to that of classical long-range Ising chain with $1/r^{2-\delta}$ interaction\cite{fisher1972critical,luijten1997classical}, thereby a quantum-classical correspondence is established\cite{guo2012critical}.
However, introducing additional spin components  coupling to the bath can significantly change the phase diagram of the system. A non-classical phase has been predicted for two-component case and thus the quantum-classical correspondence is violated\cite{guo2012critical,bruognolo2014two}. The SU(2) three-component case (or Bose-Kondo model) has also been widely investigated\cite{sengupta2000spin,smith1999non,sachdev1993gapless,sachdev2004quantum,sachdev1999quantum,vojta2000quantum,si2001locally,zarand2002quantum,zhu2002critical,vojta2003pseudogap,zhu2004quantum,novais2005frustration,pixley2013quantum,chowdhury2021sachdev,joshi2020deconfined} owe to its relevance to various problems in condensed matter physics such as magnetic moments in quantum critical magnets, Kondo breakdown transitions in heavy-fermion metals and cavity quantum electrodynamics\cite{weber20222}.  Recently, more attention on the tunable behavior in such systems has been focused by both analytical\cite{nahum2022fixed,cuomo2022spin,beccaria2022wilson} and numerical\cite{weber20222} studies.

Fixed point collision and annihilation (FPCAA) is a phenomenon in renormalization group (RG) which states that an unstable fixed point collides and then annihilates with a stable fixed point when a parameter not flowing in RG process, is continously tuned. The RG flow becomes very slow in the vicinity just beyond the annihilation point and thus can lead to   interesting concepts including quasiuniversality that relates to anomalously small masses of particle and  weakly first-order phase transitions with extremely
long correlation lengths\cite{nienhuis1979first,zumbach1993almost,wang2017deconfined}.   FPCAA typically exists in the SU(2)-symmetric Bose-Kondo model where $\delta$,  i.e. the exponent of spectral density $\omega^{1-\delta}$ of the bath can be altered to make the critical point annihilate with a stable fixed point corresponding to a non-classical phase. Using the background field RG method, Ref.(\cite{nahum2022fixed}) predicted that for $0<\delta<\delta_c$ ($\delta_c=\frac{1}{\pi S}$) there exist both of the two fixed points. At $\delta_c$ the two fixed points collide and thereafter disappear when $\delta$ is beyond $\delta_c$.  Ref.(\cite{ weber20222}) confirmed the collision and annihilation phenomenon of the two intermediate-coupling RG fixed points from high-accuracy quantum Monte Carlo calculations. (In Ref.(\cite{ weber20222}) they use $s=1-\delta$ instead of $\delta$ as the tunable parameter) However, the tunability of $\delta$ may need more effort to be realized from physical consideration\cite{nahum2022fixed}.  We herein proposed a dissipative two-spin model where the FPCAA can be easily achieved by directly adjusting the distance between the two spins, thus creating an alternative to realization of such continuous tunability of phase transition.  
\section { model}
\begin{figure}
\centering
\includegraphics[width=0.42\textwidth]{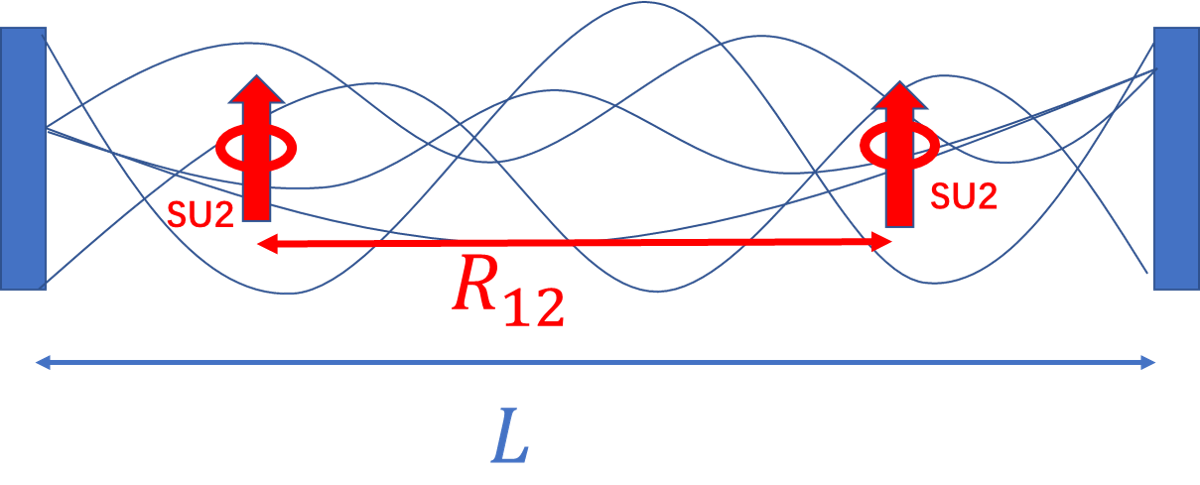}
\caption{Schematic diagram of two spins at  a distance $R_{12}$  coupled to a one-dimensional common bosonic bath with size $L$.  The coupling is SU(2)-symmetric.}\label{Fig.1}
\end{figure}

We consider two-spin generalization of Bose-Kondo model where the spin-bath coupling is also SU(2) symmetric. Our quantum model is described by
\begin{small}
\begin{align}
\mathcal{H}=\sum_{q}\omega_q \vec a_q^{\dagger}\vec a_q
+\sum_{i=1}^2 \sum_{q}\lambda_q(\vec a_q^{\dagger}e^{i\vec q \cdot \vec r_i }+\vec a_q e^{-i\vec q \cdot \vec r_i})\cdot \vec S_i \label{1}
\end{align}
\end{small} 
Where two individual spins located at the positions $\vec r_i$ are coupled to a common bath whose spectral is assumed to be  $J(\omega)=\alpha\Lambda^\delta \lvert \omega\rvert^{1-\delta}$. $\alpha$ is the dissipation strength. $\Lambda$ is the frequency cutoff. Note that the coupling here differs from that of the dissipative spin chain\cite{werner2005phase,schehr2006strong,hoyos2008theory,hoyos2012dissipation,weber2022dissipation,werner2005quantum}, where each spin is embedded in its own bosonic bath (known as the site dissipation). Some early works\cite{dube1998dynamics,mccutcheon2010separation,winter2014quantum} addressed similar two-spin models but the spin-bath coupling is single-component case instead of SU(2), just as single-component spin-boson model discussed at the beginning of the paper. The phase factor $e^{-i\vec q \cdot \vec r_i}$ in Eq.(\ref{1}) accounts for the position-dependent coupling between the spins and the bath. This model could be implemented, for instance, by placing two spins in a box with a linear size $L$ where the bosonic modes are represented by standing waves  with periodic boundary conditions (see Fig.(\ref{Fig.1})).
\\

After integrating out the bath degree of freedom (See Supplemental Material\cite{supplement} for detail of  path integral of the quantum model), an effective action of this problem manifests itself as
\begin{align} 
\mathcal{S}=\Omega[\vec n_1,\vec n_2]+{S}_{ret}\label{2}
\end{align} 
Where the non-local (or long-ranged) action is
\begin{small}
\begin{align}
{S}_{ret}=\frac{1}{2g}\sum_{ij}\iint dt dt^{\prime}  \widetilde{K}(t-t^{\prime},\vec r_i-\vec r_j)[\vec n_i(t)-\vec n_j(t^{\prime})]^2
\end{align}
\end{small}
Where $g^{-1}=\alpha S^2$. $\Omega[\vec n_1,\vec n_2]=iS\sum_{i=1}^{2}\int dt \partial_t \phi_{i}(1-\mathrm {cos}\theta_i)$ is the Berry term. The normalized spin fields $\vec n_1$ and $\vec n_2$ have been parameterized as $\vec n_i (t)= (\mathrm {sin}\theta_i \mathrm {cos}\phi_i,\mathrm {sin}\theta_i \mathrm {sin}\phi_i,\mathrm {cos}\theta_i)$.
Compared to the model in Ref.(\cite{nahum2022fixed}), ${S}_{ret}$ contains not only a retarded interaction in time but an effective ferromagnetic interaction between the two spins despite    the absence of a direct spin-spin interaction in the original quantum model. The integral kernel $K(t-t^{\prime},\vec R_{ij})=\alpha \widetilde{K}(t-t^{\prime},\vec R_{ij})$ is
\begin{align}
K(t, \vec R_{ij})=\int _0^{\Lambda}J(\omega,\vec R_{ij})\frac{\mathrm {cosh}[\omega(\beta/2-t)]}{2\mathrm {sinh(\beta \omega/2)}}d\omega
\end{align}
Where the dependence on  $\vec R_{ij}=\vec r_i-\vec r_j$ can be completely integrated into the spectral function $J(\omega,\vec R_{ij})=\alpha\Lambda^\delta \lvert \omega\rvert^{1-\delta}f^d(\widetilde R_{ij})$, $f^d(\widetilde R_{ij})$ is a function of $R_{ij}$ whose value is between $-1\sim 1$ (See Ref(\cite{winter2014quantum}) and also Supplemental Material\cite{supplement} ).   $\widetilde R_{ij}=\frac{R_{ij}\Lambda}{v}$ and $ R_{ij}=\lvert \vec r_i-\vec r_j \rvert$. $v$ is the propagation velocity for the information  propagating from one spin to the other spin, and it can be explained by the dispersion of the bosonic field, i.e. $\omega_{\vec q}=v\lvert \vec q \rvert$. 
In the next section, we will study the action\ref{2} using a Background field RG method firstly used in Ref.(\cite{nahum2022fixed}).   
\section {RG results and discussion}
Our main result is the RG equation
\begin{align}
\frac{dh}{dl}=-\delta h + h\frac{2}{\pi S}\frac{h(1-x^2)+h^3}{h^4+2h^2(1+x^2)+(x^2-1)^2}\label{5}
\end{align}
Where $h=gS$ and $x=f^d(\widetilde R_{12})$. Now we have two parameters, i.e. $\delta$ and $x$, affecting the  behavior of the beta function. Through altering the distance between the two spins, $x$ can be continously tuned as well. Thus we expected FPCAA could be achieved by simply changing the distance. Before  showing this, we will firstly indentify typical fixed points which may present in this model.\\
\begin{figure}
\centering
\includegraphics[width=0.4\textwidth]{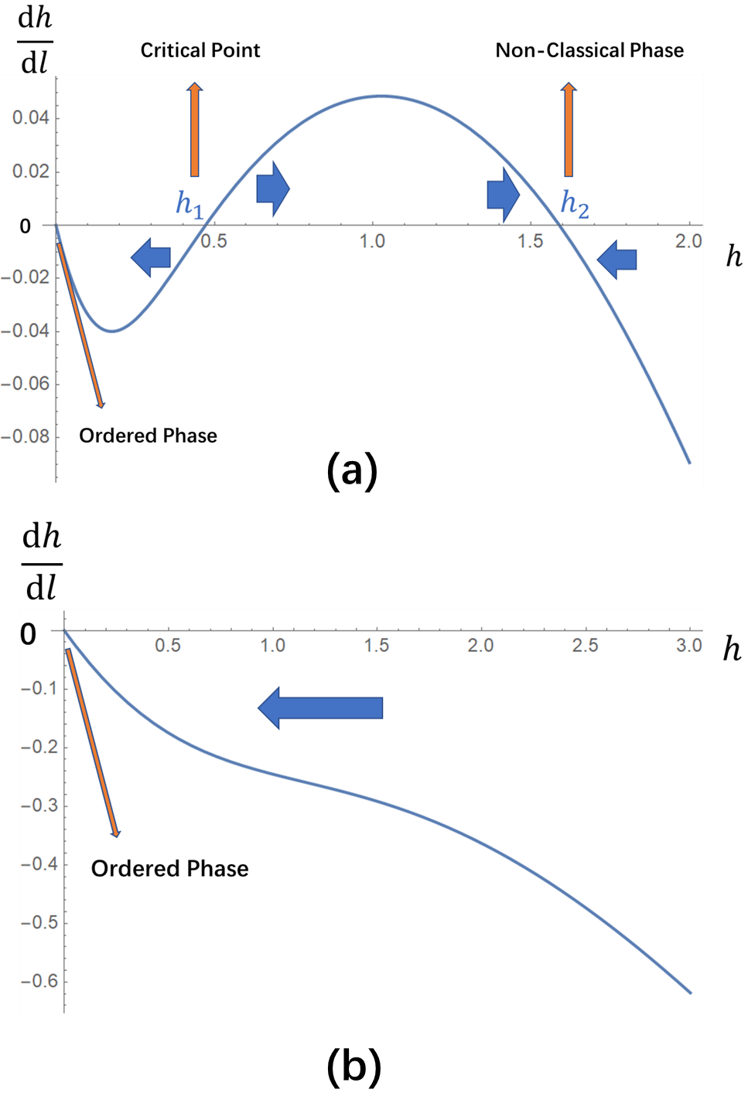}
\caption{\textbf{Plot of the calculated beta function.}   In (a) $h_1$ is the phase transition point and $h_2$ is an stable fixed point corresponding to a non-classical phase. $h=0$ is a stable fixed point representing the ordered phase and $h=\infty$ indicates an unstable fixed point (not shown). In (b) all flows lead to the ordered phase. Here $x=0.5$ in (a) and $x=1$ in (b).  $\delta=0.5$, $S=1/2$ in both (a) and (b).} \label{Fig.2}
\end{figure}\\
\begin{figure*}
\centering
\includegraphics[width=0.8\textwidth]{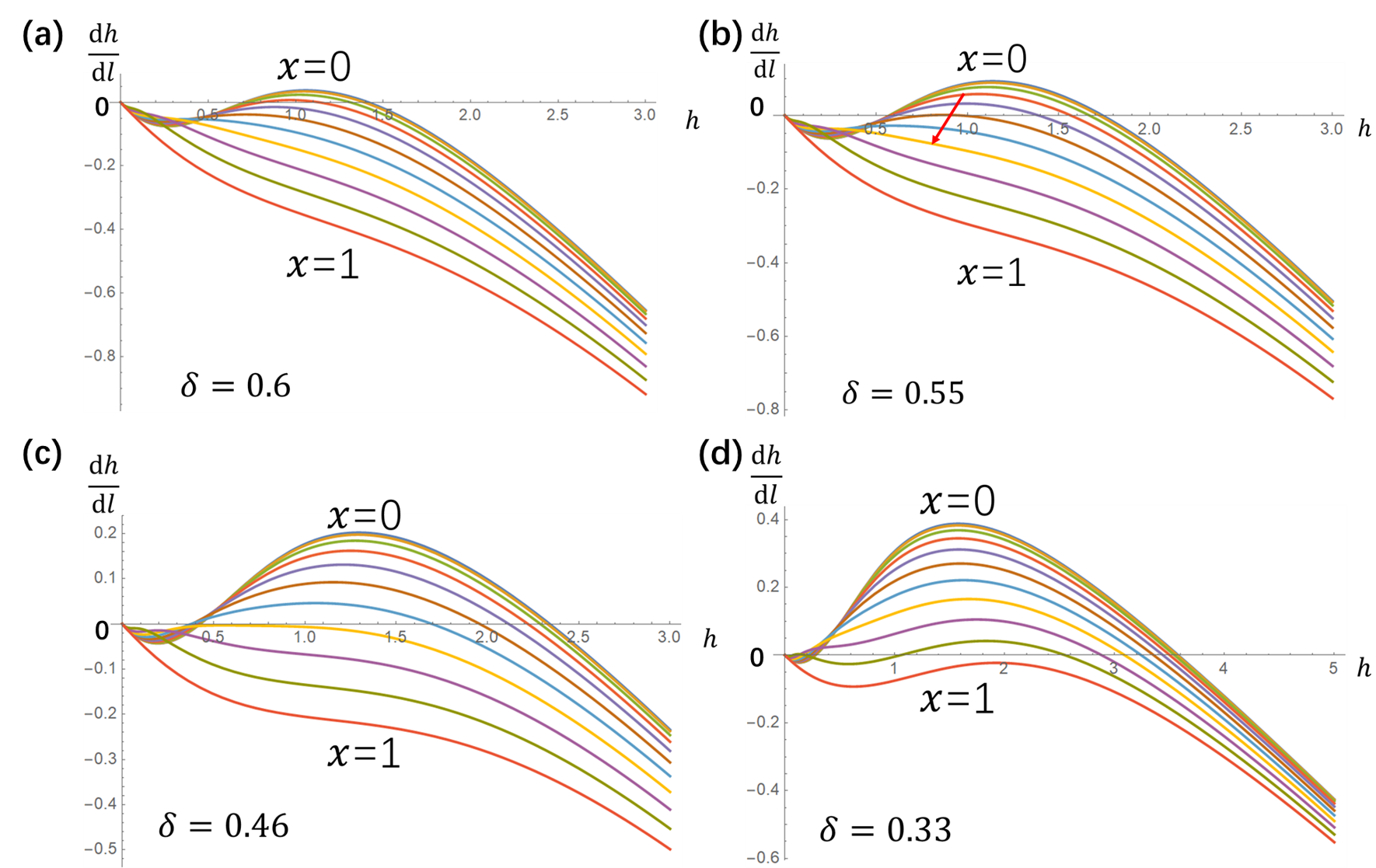}
\caption{\textbf{Fixed points collision and annihilation by adjusting the distance between two spins.} In each diagram, $x$ is changed from $0\sim 1$ with an interval 0.1, and the corresponding curves of the beta function are ploted from top to bottom. The rea arrow in (b) marks occurance of a FPCAA. $\delta$ is chosed to be 0.6, 0.55, 0.46 and 0.33 from (a) to (d), respectively. $S=1/2$   }\label{Fig.3}
\end{figure*}
Fig.(\ref{Fig.2} a) is a plot of the beta function with parameters $\delta=0.5$, $x=0.5$. The situation is quite similar to that in Ref(\cite{nahum2022fixed}). We can see there exist both a phase transition point $h_1$, and a stable phase point $h_2$ representing a non-classical phase with finite coupling, which do not appear in the corresponding classical long-ranged spin  model without a Berry term\cite{kosterlitz1976phase}.  We calculated the spin autocorrelator $\langle \vec n_i(t)\cdot \vec n_j(t^{\prime})\rangle$ at large $\vert t-t^{\prime}\vert$. It obeys a power law decay as $\vert t-t^{\prime}\vert^{-\delta}$ at both fixed points, just like the single spin case in Ref.(\cite{nahum2022fixed}). This exponent value is expected to be exact, as for other long range models\cite{zhu2002critical,kosterlitz1976phase,fisher1972critical,brezin1976critical,paulos2016conformal}.
$h=0$ is the completely ordered phase, or strong coupling phase (Note $h\propto \alpha^{-1}$, $\alpha$ is the dissipation strength). What is worth noticing is that when $x$ is changed to 1, as shown in Fig.(\ref{Fig.2} b), the system will always flow to the strong coupling phase, no matter how weak the bare coupling is. This is in stark contrast to the classical long-ranged spin model where the system will flow to high temperature phase when the bare temperature is high enough\cite{kosterlitz1976phase}.

Now we will focus on the FPCAA in this model. Firstly we defined the case $x=0$ as Single-Spin Limit (SSL), which recreats exactly the same situation in Ref.(\cite{nahum2022fixed}). At SSL, the $\delta_c$ at which two intermediate-coupling RG fixed points $h_1$ and $h_2$ collide is $\frac{1}{\pi S}$. While at Two-Spin Limit (TSL) defined as the case $x=1$, the $\delta_c$ is $\frac{1}{2\pi S}$. Through altering the distance between the two spins in our model, we can continously changing $x$, thereby an infinite number of intermediate states between SSL and TSL can be reached.  If we set the $\delta$ in the model between $\frac{1}{2\pi S}\sim \frac{1}{\pi S}$, a FPCAA must be observed in the process of adjusting the distance. Fig.(\ref{Fig.3}) shows the FPCAA of a $S=1/2$ system from which we can see our scenario works pretty well. For different $\delta$, we have indeed observed that two untrivial fixed points annihilate at some value of x, donoted as $x_c$ in the following, after which both $h_1$ and $h_2$ become imaginary. (see Supplemental Material for the value of $x_c$ and more details) 
 \\\\
\begin{figure*}
\centering
\includegraphics[width=0.9\textwidth]{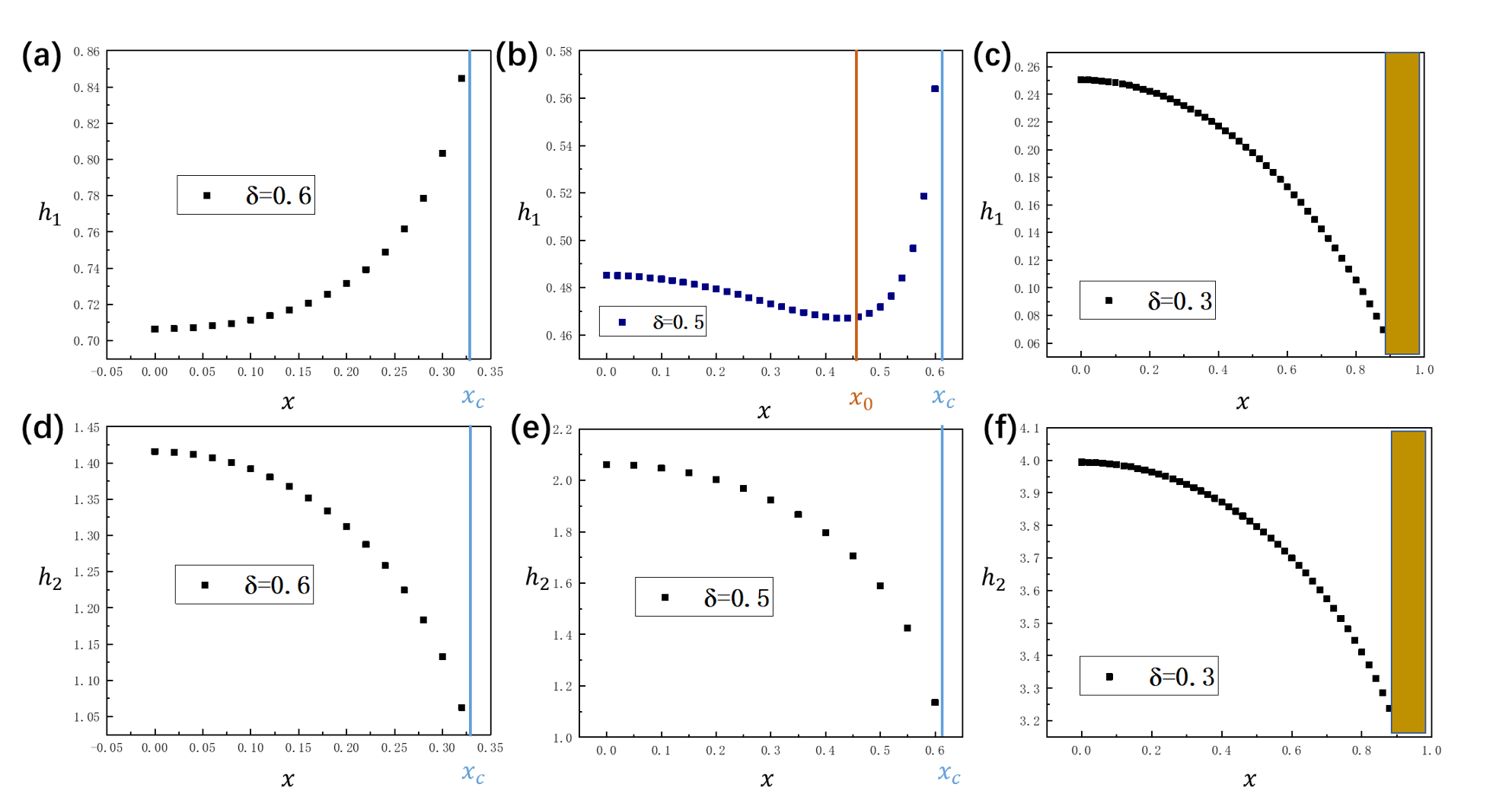}
\caption{\textbf{ Variation of the critical point $h_1$ and the stable fixed point $h_2$ by adjusting the distance.} $x$ is a function of the distance $R_{12}$ (see Eq.(\ref{5})). (a),(b) and (c) are variation of $h_1$ with $\delta=0.6$, $0.5$, $0.3$, respectively. (d),(e) and (f) are variation of $h_2$ with  same parameter as former. $x_c$ is the FPCAA point where both $h_1$ and $h_2$ become imaginary. $x_0$ is the minimal of $h_1$ in (b).  The shadow area is where the beta function oscillates and thus it is denoted as undetermined region.   }\label{Fig.4}
\end{figure*}
It is also desirable to investigate variation of the critical value $h_1$ when adjusting the distance and we still focused on the $S=1/2$ case. $h_1$ is the transition point which indicates the  minimum value of the dissipation strength $\alpha$ needed to reach the ordered phase (see Fig.(\ref{Fig.2} a)).  In fact, the variation of $h_1$ shows strong tunability with the increasing of $x$.  If $\delta$ is in the range of $\frac{1}{\pi }$ to $\frac{\sqrt{3}}{\pi }$, e.g. Fig.\ref{Fig.4} (b), the critical point $h_1$ will experience a decreasing before the increasing with the increase of $x$, but terminate at $x_c$ after which both $h_1$ and $h_2$ become imaginary.  $x_c$ is exactly where FPCAA occurs. In the range $0<\delta<\frac{1}{\pi}$, e.g. Fig.\ref{Fig.4}(c),  $h_1$ will decrease with $x$ increasing from  $0$ to $1$, while in the range $\frac{\sqrt{3}}{\pi }<\delta<\frac{2}{\pi }$, e.g. Fig.\ref{Fig.4}(a), $h_1$ will increase and terminate at $x_c$. When $\delta>\frac{2}{\pi }$, there is no untrivial fixed point no matter how $x$ changes. Fig.\ref{Fig.4} (d), (e), (f) are the corresponding variation of the stable fixed point $h_2$. Please note that the shadow area in Fig.\ref{Fig.4}(c) which begins at  $\frac{1}{2} \sqrt{4-\pi ^2 \delta^2}$ and ends at 1, is donoted as undermined region where the beta function oscillates due to the emergence of additional two fixed points. (see Supplemental Material for detail)  The new fixed points maybe an indicator of a new phase but the physical origin of them is unknown to us, thus needs further investigation. 

\section{Conclusion and Outlook}
In summary, we have proposed a SU(2) symmetric two-spin-boson model where the two spins share a commom bath. The model exhibits continous tunability where the FPCAA phenomenon could be easily achieved by simply changing the distance between the two spins. We analytically studied the corresponding field theory of the system using a Background field RG method and indeed observed FPCAA in the process of changing the distance. For $S=1/2$ spin, our scheme works pretty well for any value of $\delta$ between $\frac{1}{\pi}\sim\frac{2}{\pi}$, thus avoiding the  difficulty of fine adjusting $\delta$ in the vicinity of the annihilation point. Moreover, we also studied the variation of the positions of both the critical point and the stable fixed point. The critical value $h_1$ has a minimum if $\delta$ is in the range of $\frac{1}{\pi }$ to $\frac{\sqrt{3}}{\pi }$. These results  facilitate physical realization of FPCAA and are intended to provide some valuable suggestions of  tunable phase transition for other systems.\\
 
A possible generalization of our model is to introduce a direct interaction between the spins which may change the
universality class of the transition. Other generalizations may include $N$ free spins or a spin chain embedded in a common bath with SU(2) spin-bath coupling. It will be worthwhile to study these variety of models using RG method, as well as numerical techniques such as Monte Carlo simulations with a continuous imaginary time cluster algorithm, which is used in Ref(\cite{winter2014quantum})  or high-accuracy quantum Monte Carlo calculations for retarded interactions used in Ref(\cite{weber20222}).

\section*{Acknowledgement}
I thank Zi Cai for useful discussions, and also Mingxi Yue, Shuohang Wu, and Chenyue Guo for suggestions on the manuscript. This work is supported by the National Key Research and Development Program of China (Grant No. 2020YFA0309000), NSFC of  China (Grant No.12174251), Natural Science Foundation of Shanghai (Grant No.22ZR142830),  Shanghai Municipal Science and Technology Major Project (Grant No.2019SHZDZX01).
\bibliography{Tunable.bib}

\section{Supplemental Material for \\
Tunable phase transition in a dissipative two-spin system: A reormalization group study}

\textbf{Path integral of the quantum model} - After putting the Hamiltonian (Eq.(\ref{1}) in the main text) into the machinery of both spin-path integral and the path integral for bosons, the partition function  could be written as 
\begin{align}
Z=\int \mathnormal{D} \vec n_i(t) \delta(\vec n_i^2-1)\mathnormal{D}(\vec a_q^{\ast},\vec a_q) e^{-\mathcal{S}}
\end{align}
\begin{align}
\mathcal{S}=\Omega[\vec n_1,\vec n_2]+\mathcal{S}_b+\mathcal{S}_{sb}
\end{align}
\begin{align}
\Omega[\vec n_1,\vec n_2]=iS\sum_{i=1}^{2}\int dt \partial_t \phi_{i}(1-\mathrm {cos}\theta_i)
\end{align}
We have parameterized $\vec n_i (t)$ as \\$\vec n_i (t)= (\mathrm {sin}\theta_i \mathrm {cos}\phi_i,\mathrm {sin}\theta_i \mathrm {sin}\phi_i,\mathrm {cos}\theta_i)$.
\begin{align}
\mathcal{S}_b=\int dt \sum_{q} \vec a_q^{\ast}(t)(\partial_t+\omega_q)\vec a_q(t)
\end{align}
\begin{small}
\begin{align}
\mathcal{S}_{sb}=S\int dt \sum_{i=1}^2 \sum_{q}\lambda_q[\vec a_q^{\ast}(t)e^{i\vec q \cdot \vec r_i }+\vec a_q(t) e^{-i\vec q \cdot \vec r_i}]\cdot \vec n_i(t) 
\end{align}
\end{small}
Now we integrate out the bath degree. A retarded interaction in time appears, namely,
\begin{small}
\begin{align}
\mathcal{S}_{ret}=-S^2\sum_{ij}\iint dt dt^{\prime}  K(t-t^{\prime},\vec r_i-\vec r_j)\vec n_i(t)\cdot \vec n_j(t^{\prime})\label{9}
\end{align}
\end{small}
Intriguingly, now the two individual spins have an effective ferromagnetic interaction in $\mathcal{S}_{ret}$ despite they are independent in the original Hamiltonian.
The integral kernel $K(t-t^{\prime},\vec r_i-\vec r_j)$ is
\begin{align}
K(t, \vec R_{ij})=\int _0^{\Lambda}J(\omega,\vec R_{ij})\frac{\mathrm {cosh}[\omega(\beta/2-t)]}{2\mathrm {sinh(\beta \omega/2)}}d\omega
\end{align}
Where  $J(\omega,\vec R_{ij})=\sum_{\vec q>0}\lambda_{\vec q}^2    \mathrm{cos}(\vec q \cdot \vec R_{ij})\delta(\omega-\omega_{\vec q})$. After assuming linear dispersion of bath, i.e. $\omega_{\vec q}=v\lvert q \rvert$,  $J(\omega,\vec R_{ij})=\alpha\Lambda^\delta \lvert \omega\rvert^{1-\delta}f^d(\widetilde R_{ij})$, where $\widetilde R_{ij}=\frac{R_{ij}\Lambda}{v}$, $ R_{ij}=\lvert \vec r_i-\vec r_j \rvert$. 
 In different dimensions, $f^d(\widetilde R_{ij})$ has different forms. $f^d(\widetilde R_{ij})=\mathrm{cos}(\widetilde R_{ij})$, $J_0(\widetilde R_{ij})$, and  $\mathrm{sin}(\widetilde R_{ij})/\widetilde R_{ij}$ for $d=1,2$ and $3$, respectively\cite{winter2014quantum}. Note that $f^d(\widetilde R_{ij})$ is in the interval between $-1\sim 1$ for all dimensions.
Now let $ K(t-t^{\prime},\vec R_{ij})=\alpha \widetilde{K}(t-t^{\prime},\vec R_{ij})$. Then Eq.(\ref{9}) can be rewritten as 
\begin{small}
\begin{align}
\mathcal{S}_{ret}=\frac{1}{2g}\sum_{ij}\iint dt dt^{\prime}  \widetilde{K}(t-t^{\prime},\vec r_i-\vec r_j)[\vec n_i(t)-\vec n_j(t^{\prime})]^2
\end{align}  
\end{small}
 Note that the dissipation strength $\alpha$ has already been absorbed into the defined coupling constant $g^{-1}=\alpha S^2$.\\

\textbf{Background field renormalization} - We used a background field method in \cite{nahum2022fixed}to do the renormalization.

Firstly, we introduced a slowly varying background field
\begin{align}
\vec n_{is}(t)=(\mathrm {cos} \phi_{is}(t),\mathrm{sin}\phi_{is}(t),0)
\end{align}
and the full field is
\begin{small}
\begin{align}
\vec n_i(t)=(\sqrt{1-\chi_{if}^2}[\mathrm {cos} (\phi_{is}+\phi_{if}),\mathrm {sin} (\phi_{is}+\phi_{if})],\chi_{if})
\end{align}
\end{small}
$i=1,2$.\\
We would express the action in terms of $(\chi_{is}$,$\chi_{if}$,$\phi_{is}$,$\phi_{if}$), up to quadratic order in the fast fields.
\begin{small}
\begin{align}
\Omega[\vec n_1,\vec n_2]\simeq\Omega[\vec n_{1s},\vec n_{2s}]-iS\sum_{i=1}^{2}\int dt \chi_{if}\partial_t \phi_{if}
\end{align}
\end{small}
The Berry term will not get renormalized because $\Omega[\vec n_{1s},\vec n_{2s}]$ does not contain fast fields.

\begin{small}
\begin{align}
[\vec n_i(t)-\vec n_j(t^{\prime})]^2\simeq(\chi_{if}-\chi_{jf}^{\prime})^2+(\phi_{if}-\phi_{jf}^{\prime})^2\nonumber\\
+[\vec n_{is}(t)-\vec n_{js}(t^{\prime})]^2(1-\frac{1}{2}[\chi_{if}^2+\chi_{jf}^{\prime2}+(\phi_{if}-\phi_{jf}^{\prime})^2])
\end{align}
\end{small}
Where $\chi_f^{\prime}=\chi_f(t^{\prime})$,$\phi_f^{\prime}=\phi_f(t^{\prime})$,$\vec n_s^{\prime}=\vec n_s(t^{\prime})$.

Now we put altogether all of the fast parts of the action as the Gaussian term under which the pertubation will be implemented.

\begin{small}
\begin{align}
\mathcal{S}_{f}=
-iS\sum_{i=1}^{2}\int dt \chi_{if}\partial_t \phi_{if}
\nonumber \\+\frac{1}{2g}\sum_{ij}\iint dt dt^{\prime}  \widetilde{K}(t-t^{\prime},\vec R_{ij})\nonumber\\
[(\chi_{if}-\chi_{jf}^{\prime})^2+(\phi_{if}-\phi_{jf}^{\prime})^2]
\end{align}
\end{small}
And the rest is
\begin{small}
\begin{align}
\mathcal{S}_{fs}=
\frac{1}{2g}\sum_{ij}\iint dt dt^{\prime}  \widetilde{K}(t-t^{\prime},\vec R_{ij})[\vec n_{is}(t)-\vec n_{js}(t^{\prime})]^2
\nonumber\\
(1-\frac{1}{2}[\chi_{if}^2+\chi_{jf}^{\prime2}+(\phi_{if}-\phi_{jf}^{\prime})^2])\label{10}
\end{align}
\end{small}
Now we are in a position to do the renormalization. Firstly we  represent $\mathcal{S}_{f}$ in Fourier space. From the spectrum function $J(\omega,\vec R_{ij})$, the Fourier transformation to frequency space of $\widetilde{K}(t-t^{\prime},\vec R_{ij})$ can be evaluated as $\Lambda^\delta \lvert \omega\rvert^{1-\delta}f^d(\widetilde R_{ij})$. Then
\begin{align}
\mathcal{S}_{f}=\frac{1}{2}\int \frac{d\omega}{2\pi}\Psi^T(\omega)G\Psi(-\omega)
\end{align}
Where $\Psi(\omega)=(\chi_{1f}(\omega),\phi_{1f}(\omega),\chi_{2f}(\omega),\phi_{2f}(\omega))^T$
\begin{widetext}
\[
  G = 
  \begin{pmatrix}
    g^{-1}\Lambda^\delta \lvert \omega\rvert^{1-\delta} & S\omega & g^{-1}\Lambda^\delta \lvert \omega\rvert^{1-\delta}f^d(\widetilde R_{12}) & 0\\
   -S\omega & g^{-1}\Lambda^\delta \lvert \omega\rvert^{1-\delta} & 0 &g^{-1}\Lambda^\delta \lvert \omega\rvert^{1-\delta}f^d(\widetilde R_{12})\\
g^{-1}\Lambda^\delta \lvert \omega\rvert^{1-\delta}f^d(\widetilde R_{12})& 0 & g^{-1}\Lambda^\delta \lvert \omega\rvert^{1-\delta}&S\omega\\
0 & g^{-1}\Lambda^\delta \lvert \omega\rvert^{1-\delta}f^d(\widetilde R_{12})& -S\omega& g^{-1}\Lambda^\delta \lvert \omega\rvert^{1-\delta}
  \end{pmatrix}
\]
\end{widetext}
 From the expression of $G$, we can obtain the Gaussian average under $\mathcal{S}_{f}$
\begin{small}
\begin{align}
\langle \mathcal \chi_{if}^2 \rangle=\langle \mathcal \phi_{if}^2 \rangle=2\int_{\Lambda/b}^{\Lambda}\frac{d\omega}{2\pi}\frac{1}{\Lambda S}\frac{h(1-x^2)+h^3}{h^4+2h^2(1+x^2)+(x^2-1)^2}\nonumber\\
=\frac{l}{\pi S}\frac{h(1-x^2)+h^3}{h^4+2h^2(1+x^2)+(x^2-1)^2}\label{12}
\end{align}
\end{small}
for $i=1,2$.
Where $l=1-\frac{1}{b}\simeq \mathrm{lnb}$,   $x=f^d(\widetilde R_{12})$. We have already defined $h=gS$ since spin size $S$ does not get renormalized in the RG process.

We could estimate the RG correction using
\begin{align}
\mathit{Z}=\int \mathnormal{D}\phi_s\mathnormal{D}(\phi_f,\chi_f)e^{-\mathcal{S}_{f}}e^{-\mathcal{S}_{fs}}\nonumber\\
\simeq \int \mathnormal{D}\phi_s e^{-\langle \mathcal{S}_{fs}\rangle +..}
\end{align}

According to Eq.(\ref{10}) and Eq.(\ref{12}), the coupling constant runs as
\begin{align}
\frac{1}{2g}\to \frac{1}{2g}b^{\delta}(1-[\langle \chi_{if}^2 \rangle +\langle \phi_{if}^2 \rangle ])\label{21}
\end{align}
Where we have dropped the two-point function of the fast field at distinct imaginary time, i.e $\langle \phi_{if}\phi_{jf}^{\prime} \rangle $. The term $b^{\delta}$ comes from rescale of the imaginary time $t\to bt$. It can be seen from Eq.(\ref{12}) that both $\langle \mathcal \chi_{if}^2 \rangle$ and $\langle \mathcal \phi_{if}^2 \rangle$   are order of $\frac{1}{S}$. Thus the correction to the coupling constant in Eq.(\ref{21}) is also order of $\frac{1}{S}$. It implies our calculation seems valid at large $S$ limit but in fact it also applies for all values of the spin\cite{nahum2022fixed}, including
$S=1/2$, since the topology of the flows remain unchanged.
Then 
\begin{align}
\frac{dh}{dl}=-\delta h + h\frac{2}{\pi S}\frac{h(1-x^2)+h^3}{h^4+2h^2(1+x^2)+(x^2-1)^2}
\end{align}

For sake of the spin autocorrelator $\langle \vec n_i(t)\cdot \vec n_j(t^{\prime})\rangle$ (Note: unlike $\langle \ \rangle$ donoted as Gaussian average under fast fields in the previous RG calculation, here it means the average under the full action) at the nontrivial fixed points. We firstly find the scaling dimension $z$ of the field $\vec n_i$. We assumed $\langle \vec n_i \rangle_f =b^{-z} \vec n_{is}$ under rescaling of the imaginary time. $\langle \ \rangle_f$ is donoted as the Gaussian average under the fast fields for clarification. We adopted $\vec n_{is}=(1,0,0)$, then
\begin{align}
\langle \vec n_i \rangle_f=(1-\frac{1}{2}[\langle \chi_{if}^2 \rangle_f +\langle \phi_{if}^2 \rangle_f ])\vec n_{is}
\end{align}
and 
\begin{align}
z=\frac{1}{\pi S}\frac{h(1-x^2)+h^3}{h^4+2h^2(1+x^2)+(x^2-1)^2}
\end{align}  
In this way the RG equation has a form $\frac{dh}{dl}=-\delta h + 2zh$. Thus at both the two untrivial fixed points $z=\frac{\delta}{2}$.  So when $\vert t-t^{\prime}\vert$ is large, the spin autocorrelator $\langle \vec n_i(t)\cdot \vec n_j(t^{\prime})\rangle$ has a power law decay as $\vert t-t^{\prime}\vert^{-\delta}$  at both fixed points, just like the single spin case in Ref.(\cite{nahum2022fixed}).\\

\textbf{ Roots of the beta function } - The parameters space considered is $x\in 0\sim1$ and $\delta\in 0\sim 1$. We focused on the case of $S=1/2$. The four roots of the beta function are 
\\
\begin{small}
\begin{align}
a_1=A-\frac{\sqrt{B+C+\frac{2}{\delta^2}-\pi ^2}}{\pi }+\frac{1}{\pi  \delta}
\end{align}
\end{small}

\begin{small}
\begin{align}
a_2=A+\frac{\sqrt{B+C+\frac{2}{\delta^2}-\pi ^2}}{\pi }+\frac{1}{\pi  \delta}
\end{align}
\end{small}\\
\begin{small}
\begin{align}
a_3=-A-\frac{\sqrt{-B-C+\frac{2}{\delta^2}-\pi ^2}}{\pi }+\frac{1}{\pi  \delta}
\end{align}
\end{small}\\
\begin{small}
\begin{align}
a_4=-A+\frac{\sqrt{-B-C+\frac{2}{\delta^2}-\pi ^2}}{\pi }+\frac{1}{\pi  \delta}
\end{align}
\end{small}\\
Where $A=\frac{\sqrt{1-\pi ^2 \delta^2 x^2}}{\pi  \delta}$, $B=-\frac{2 \pi ^2 x^2}{\sqrt{1-\pi ^2 \delta^2 x^2}}$, $C=\frac{2}{\delta^2 \sqrt{1-\pi ^2 \delta^2 x^2}}$.
The condition for $a_1$ and $a_2$ to be real can be deduced from the following equations,
\begin{align}
1-\pi ^2 \delta^2 x^2 \ge 0
\end{align} 
\begin{align}
-\frac{2 \pi ^2 x^2}{\sqrt{1-\pi ^2 \delta^2 x^2}}+\frac{2}{\delta^2 \sqrt{1-\pi ^2 \delta^2 x^2}}+\frac{2}{\delta^2}-\pi ^2\ge 0
\end{align} 
For $0<\delta<\frac{\sqrt{2}}{\pi }$, the value of $x$ should be in the range of $0\sim \mathrm{min}[ \frac{1}{\pi \delta }$,1] for $a_1$ and $a_2$ to be real. And for $\frac{\sqrt{2}}{\pi }<\delta<\frac{2}{\pi }$ , $0<x<\frac{1}{2} \sqrt{4-\pi ^2 \delta^2}$ ensures the realness of $a_1$ and $a_2$. For $\delta>\frac{2}{\pi}$, $a_1$ and $a_2$ are always imaginary. \\
The conditions for $a_3$ and $a_4$ to be real can also be deduced in the same way. In Table(\ref{table:1}) all possibilities of the solutions have been exhausted. We note that the stable fixed point $h_2$ in the main text is always equal to $a_2$ and in most cases the critical point $h_1$  is equal to $a_1$. The siuation becomes complicated only if $\delta<\frac{\sqrt{2}}{\pi}$ with $\frac{1}{2} \sqrt{4-\pi ^2 \delta^2}<x<\frac{1}{\pi \delta}$ where there exist four real roots. This can change the behavior of FPCAA.  We will discuss it in detail in the following section.\\
\begin{table}[htb]   
\begin{center}   
\caption{Realness of the solutions.}  
\label{table:1} 
\begin{tabular}{|m{3.2cm}<{\centering}|m{1.2cm}<{\centering}|m{1.2cm}<{\centering}|m{1.2cm}<{\centering}|m{1.2cm}<{\centering}|}   
\hline   \textbf{} & \textbf{$a_1$} & \textbf{$a_2$}& \textbf{$a_3$}& \textbf{$a_4$}   \\   
\hline   $\frac{2}{\pi}<\delta<1$ & imag & imag & imag& imag \\ 
\hline   $\frac{\sqrt{2}}{\pi}<\delta<\frac{2}{\pi}$ and $\frac{1}{2} \sqrt{4-\pi ^2 \delta^2}<x<1$ & imag & imag& imag & imag \\ 
\hline   $\frac{\sqrt{2}}{\pi}<\delta<\frac{2}{\pi}$ and $0<x<\frac{1}{2} \sqrt{4-\pi ^2 \delta^2}$ & \textbf{real} & \textbf{real}& imag & imag \\  
\hline  $0<\delta<\frac{\sqrt{2}}{\pi}$ and $0<x<\frac{1}{2} \sqrt{4-\pi ^2 \delta^2}$ & \textbf{real}& \textbf{real}& imag& imag \\      
\hline 
$0<\delta<\frac{\sqrt{2}}{\pi}$ and $\frac{1}{2} \sqrt{4-\pi ^2 \delta^2}<x<\frac{1}{\pi \delta}$ & \textbf{real} & \textbf{real}& \textbf{real}& \textbf{real}\\      
\hline  
 $0<\delta<\frac{\sqrt{2}}{\pi}$ and $\frac{1}{\pi \delta}<x<1$ & imag & imag& imag& imag\\      
\hline   
\end{tabular}   
\end{center}   
\end{table}

\textbf{ The value of $x_c$ } - From the discussion in III, we can give the value of $x_c$ at which FPCAA occurs. In fact, $x_c$ should be equal to the value of $x$ after which both $h_1$ and $h_2$ become imaginary. Thus,

$x_c= \frac{1}{\pi \delta}$ for $\frac{1}{\pi}<\delta<\frac{\sqrt{2}}{\pi }$,\\\\
$x_c=\frac{1}{2} \sqrt{4-\pi ^2 \delta^2}$ for $\frac{\sqrt{2}}{\pi }<\delta<\frac{2}{\pi }$\\
 
\textbf{ FPCAA in different regions of $\delta$} - As noted in the main text, FPCAA occurs in the region $\frac{1}{\pi}<\delta<\frac{2}{\pi}$.  But from the discussion in III, we know that FPCAA may behave differently in $\frac{\sqrt{2}}{\pi}<\delta<\frac{2}{\pi}$ (which corresponds to Fig.(3a),(3b) and (3c) in the main text) and $\frac{1}{\pi}<\delta<\frac{\sqrt{2}}{\pi}$ (which corresponds to Fig.(3d) in the main text).
We take $\delta=0.55$  , i.e.Fig.(3b), and $\delta=0.33$, i.e.Fig.(3d) in the main text, as examples and analyze them in detail the behavior of the beta function in the vicinity of $x_c$. For $\delta=0.55$, $x_c=0.503598$; and for $\delta=0.33$, $x_c=0.964575$.\\
\begin{figure}
\centering
\includegraphics[width=0.4\textwidth]{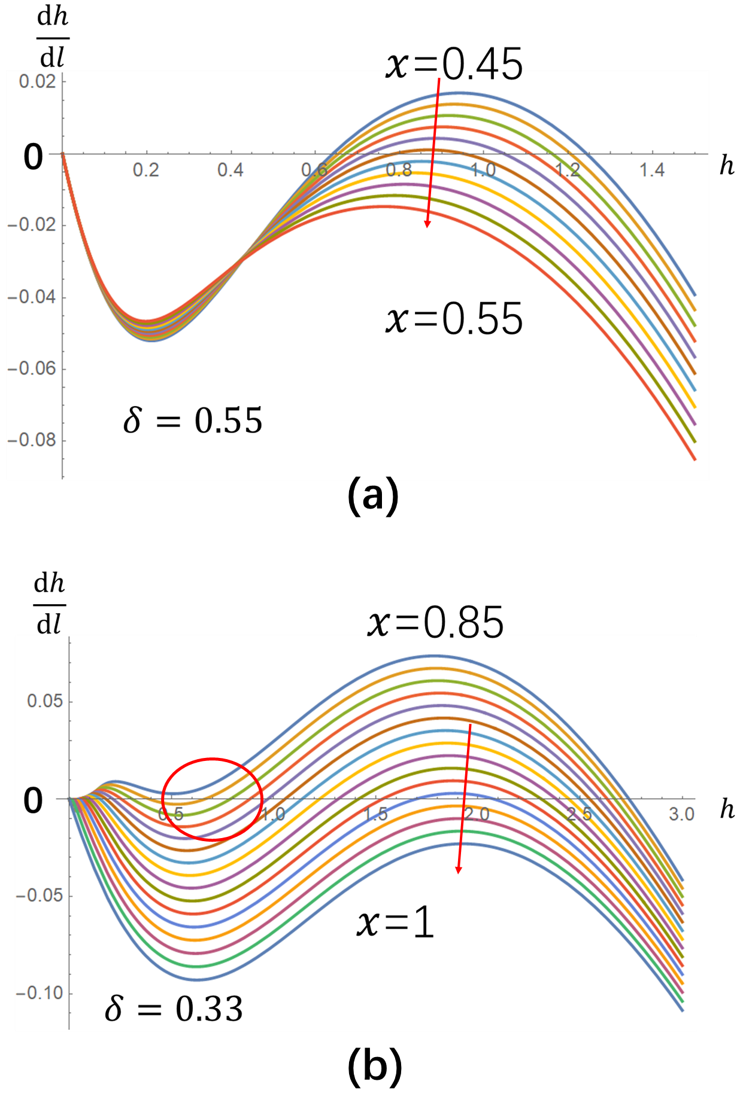}
\caption{\textbf{Fixed points collision and annihilation by adjusting the distance between two spins.} In (a), $x$ is changed from $0.45\sim 0.55$ with an interval 0.01, and the corresponding curves of the beta function are plotted from top to bottom. In (b), $x$ is changed from $0.85\sim 1$. The rea arrow  marks occurance of  FPCAA. And The red circle marks emergence of a new fixed point. $\delta$ is chosed to be 0.55, 0.33 in (a) and (b), respectively. $S=1/2$   }\label{Fig.S1}
\end{figure}\\
In Fig.(\ref{Fig.S1} a), we can see $h_1$ and $h_2$ directly collide at $x_c$, which is in excellent agreement with the FPCAA in both Ref.(\cite{nahum2022fixed} and \cite{weber20222}). But in Fig.(\ref{Fig.S1} b), a new type of FPCAA appears. A new fixed point which corresponds to $a_4$ emerges when $\frac{1}{2} \sqrt{4-\pi ^2 \delta^2}<x<\frac{1}{\pi \delta}$ (see Table(\ref{table:1}) for a reference). Then this new fixed point collide with $h_2$ at $x_c$ instead of the original $h_1$. Meanwhile, $h_1$ collide with another new fixed point which corresponds to $a_3$. All of the four real roots become imaginary after $x_c$. This new type of FPCAA involves additional fixed points which may correspond to   new critical point and new phase. Due to the limitation of our knowlegde, we will not discuss it further.\\

\begin{figure}
\centering
\includegraphics[width=0.4\textwidth]{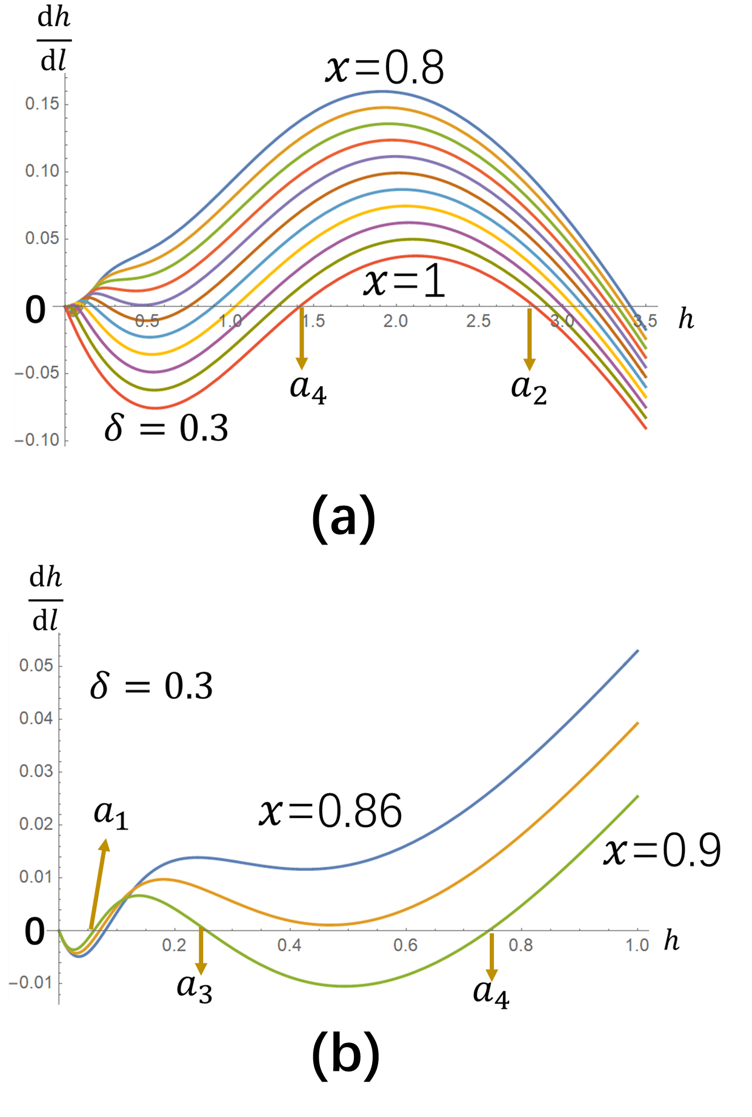}
\caption{\textbf{Oscillation of the beta function in the shadow area .} In (a), $x$ is changed from $0.8\sim 1$ with an interval 0.02, and the corresponding curves of the beta function are plotted from top to bottom. In (b), $x$ is changed from $0.86\sim 0.9$. The arrows  marks four roots of the beta function. $\delta$ is chosed to be 0.3 in (a) and (b). $S=1/2$   }\label{Fig.S2}
\end{figure}

\textbf{ Variation of the critical value $h_1$ by adjusting distance} - Let $\partial h_1 /\partial x=0$, and then the $x$ corresponding to the minimum $h_1$ is found to be
\begin{align}
x_0=\frac{\sqrt{\frac{-\pi ^4 \delta^4+2 \pi ^2 \delta^2+3}{\delta^2}}}{2 \pi }
\end{align}
under the condition $\frac{1}{\pi }<\delta<\frac{\sqrt{3}}{\pi }$. This means if $\delta$ is in this range, the critical point $h_1$ will experience a decreasing before the increasing with the increase of $x$, but terminate at $x_c$ after which both $h_1$ and $h_2$ become imaginary.  $x_c$ is exactly where FPCAA occurs. In the range $0<\delta<\frac{1}{\pi}$, $h_1$ will decrease with $x$ increasing from  $0$ to $1$, while in the range $\frac{\sqrt{3}}{\pi }<\delta<\frac{2}{\pi }$, $h_1$ will increase and terminate at $x_c=\frac{1}{2} \sqrt{4-\pi ^2 \delta^2}$. When $\delta>\frac{2}{\pi }$, there is no untrivial fixed point no matter how $x$ changes.\\\\
In the main text, we mentioned that the beta function oscillates in the shadow region which starts at $x=\frac{1}{2} \sqrt{4-\pi ^2 \delta^2}$. Here we carefully plot the beta function of $\delta=0.3$ in Fig.(\ref{Fig.S2}). We can see the two additional roots will not appear until $x=0.882006$, and then the beta function shows an oscillating character. This oscillation will continue to exist until $x=1$. At $x=1$ both $a_1$ (the original $h_1$ before $x=0.882006$) and $a_3$ shrink to zero, and $a_4$ becomes the new critical point $h_1$ of TSL defined in the main text. For $\delta=0.3$, the $h_1(TSL)=1.41273$.
\\

\end{document}